\documentclass[epj]{svjour}

\usepackage{psfig}
\usepackage{graphics}

\begin{document}

\title{Random inelasticity and velocity fluctuations
in a driven granular gas} 
 
\author{Alain Barrat
%\footnote{Electronic Address: Alain.Barrat@th.u-psud.fr}
and Emmanuel Trizac}
%\footnote{Electronic Address: Emmanuel.Trizac@th.u-psud.fr}

\institute{
 Laboratoire de Physique Th{\'e}orique
(UMR 8627 du CNRS), B{\^a}timent 210, Universit{\'e} de
Paris-Sud, 91405 Orsay Cedex, France 
}

\abstract{
We analyze the deviations from Maxwell-Boltzmann statistics found in
recent experiments studying velocity distributions in two-dimensional
granular gases driven into a non-equilibrium stationary state by a
strong vertical vibration.  We show that in its simplest version, 
the ``stochastic thermostat''
model of heated inelastic hard spheres, contrary to what has been
hitherto stated, is incompatible with the experimental data, although
predicting a reminiscent high velocity stretched exponential behavior
with an exponent $3/2$.
The experimental observations lead to refine a recently proposed
random restitution coefficient model.  Very good agreement is then
found with experimental velocity distributions within this framework,
which appears self-consistent and further provides relevant probes to
investigate the universality of the velocity statistics.
}

\maketitle

\section{Introduction}
Whereas equilibrium statistical mechanics has reached a ra\-ther
mature phase, the understanding of non-equilibrium processes is far
from complete.  In particular, granular (and thus inelastic)
gases~\cite{Jaeger} driven into a non-equi\-li\-brium steady state by
a suitable injection of energy define a stimulating research field
where theoretical predictions can be confronted against model
experiments, with the aim to understand the possible deviations from
equilibrium behavior. A good probe to quantify these deviations is the
velocity distribution of the grains, $P(v)$, which has focused
sustained attention recently, and has been shown to exhibit pronounced
differences from Maxwell-Boltzmann statistics
\cite{Olafsen,Losert,Rouyer,Kudrolli,Aranson}.  
Several authors reported a stretched exponential law
[on the whole range of velocities available, which covers
an accuracy of $5$ to $6$ orders of magnitude for $P(v)$]
\begin{equation}
P(v) \propto \exp[ -(v/v_0)^\nu ] \ ,
\label{eq:pv}
\end{equation}
with an exponent $\nu$ close to $3/2$~\cite{Losert,Rouyer,Aranson}
(here $v_0$ is the ``thermal'' r.m.s.  velocity). This behavior was
observed for the horizontal velocity components of a vertically
vibrated 2D system of steel beads in a wide range of driving
frequencies and densities~\cite{Rouyer}, but also in a three
dimensional electrostatically driven granular gas~\cite{Aranson}.

At this point, some questions naturally arise, that will be addressed
below: (i) is it possible to find a consistent model where this velocity
distribution would emerge?  (ii) what physical ingredients are
required? 

One possible approach consists in performing ``realistic'' molecular dynamics
simulations. The model of inelastic hard spheres (IHS) with 
binary momentum conserving collisions, and a ``reasonable'' 
restitution coefficient~\cite{Jaeger} provides the simplest
candidate. The
energy loss in a collision is proportional to the inelasticity
parameter $1-\alpha_0^2$ where $\alpha_0$ is the coefficient of normal
restitution ($0 < \alpha_0 \leq 1$), which in the simplest and
efficient approximation is a constant independent on the relative velocity
of colliding partners.
Such an approach has been presented
in~\cite{vibrated,Brey}, and allows to reproduce the experimental
velocity statistics with a good accuracy. The possible lack of 
universality has also been addressed in \cite{Brey}. 

Another route, which contrary to the previous numerical one 
has the merit to allow an analytical
derivation of $\nu$ in some cases \cite{twan}, 
consists in formulating an {\em effective} modeling of the
energy injection, considering idealized homogeneous systems of
inelastic hard spheres (given the experiments reported in
\cite{Rouyer}, the assumption of homogeneity is well founded, see below).
From this point of view, a simple and popular model consists in
IHS with constant inelasticity, with a {\em homogeneous} forcing 
described by a ``stochastic
thermostat''~\cite{Williams,Puglisi,twan,Pre,Montanero,Moon,PRE2,Garzo}. This
model has attracted attention, in particular because it has been shown
analytically \cite{twan} that $P(v)$ exhibits a high energy tail of the form of
Eq.~(\ref{eq:pv}) with $\nu=3/2$, independent on dimension and
restitution coefficient, in apparent agreement with the
experiments. This result holds at the mean-field level for an
homogeneous system. The above model, where an external white-noise
driving force acts on the particles and thus injects energy through
random ``kicks'' between the collisions, is therefore considered to
provide a relevant theoretical framework to quantify the non Gaussian
character of velocity distributions.

However, we shall see below that this uniformly heated model --in its
simplest version-- is unable
to reproduce the experimental data: if simulated in dimension 2 or
higher with an experimentally relevant value of the restitution
coefficient ($\alpha_0$ between $0.7$ and $1$), the obtained
distribution $P(v)$ is {\em indistinguishable from a Gaussian within
the experimental accuracy}~\footnote{
if this model is simulated in dimension $1$ with $\alpha_0 \in [0.7;1]$, 
the obtained non Gaussian behaviour at thermal velocities
corresponds to $\nu > 2$ instead of $\nu\simeq 3/2 $ in the experiments.}; 
in fact, the range of velocities for which
the high energy behavior $\exp[ -(v/v_0)^{3/2}]$ may be observed is
even beyond reach of precise numerical procedures and corresponds to a
regime where $P(v)$ is practically vanishing [lower than $10^{-6}P(0)$]. 
The predictions of this model are consequently incompatible
with the experimental velocity distributions, that show important non
Gaussian features already at thermal velocity
scale~\cite{Losert,Rouyer,Aranson}. 
We emphasize however that generalizations of the aforementioned heated model
have been proposed \cite{Cafiero_prl,Cafiero}. 
With a convenient choice of the extra parameters introduced, one may 
obtain a velocity distribution close to that measured in the experiments
(we will come back to this point in section \ref{sec:4}). However, 
in such approaches, energy is injected in between the collisions
whereas in the experiments we are interested in, the transfer of 
horizontal momentum
takes place at every inter-particle collision (see section \ref{sec:3}).
We will therefore focus on this feature for 
the 2D experiment reported
in~\cite{Rouyer} and investigate in details the collision dynamics in the
horizontal direction (section \ref{sec:2}).
The vibrated system under study there shows
important density and granular temperature gradients, especially close
to the boundaries which inject energy, but since the shaking 
is violent, there is a region where both gradients are very small
simultaneously. The velocity acquisition in~\cite{Rouyer} has been
restricted to this region, where the system, although open, may be
considered as homogeneous. In this article, we thus consider the following
question : remaining at the level of a homogeneous system, what
ingredients are required for a self-consistent {\em effective} description of
the horizontal degrees of freedom, that exhibit the stretched
exponential law (\ref{eq:pv})~?

\section{Effective restitution coefficients}
\label{sec:2}
In order to characterize the collision process in the horizontally
projected system, we have measured directly the effective 1D
restitution coefficient from the experimental data provided by
K. Feitosa and N. Menon, for a gas of stainless steel spheres (the
system investigated in~\cite{Rouyer}) but also for glass, brass and
aluminum beads~\cite{Feitosa}, which allow to sample a wide range of
nominal inelasticities.  Let us recall briefly the experimental
set-up.  The balls (diameter: $d=1.600 \pm 0.002~mm$) are confined to
a vertical, rectangular cage (32 $d$ high x 48 $d$ wide x 1.1 $d$
thick) sandwiched between two parallel plates of Plexiglas. The cage
is vibrated vertically at a frequency of $60~Hz$ and amplitudes up to
$2.4~d$, producing maximum accelerations, $\Gamma$, and velocities,
$v_{0}$, of $56~g$ and $1.45~m/s$ respectively.  The motions of the
balls are recorded with a high-speed camera which allows a location of
each ball with a precision of 0.03 $d$.  The results we discuss here
are taken in a rectangular (10~$d$ x 20~$d$) window around the
geometrical center of the cell, where, as mentioned above, density and
granular temperature are almost homogeneous~\cite{Rouyer}. Moreover,
the measured velocity distributions do not vary with height nor with
the phase of the vibration cycle. The experimental data can thus be
considered as obtained in the bulk of a two-dimensional homogeneous
(but open) system, reasonably far from the boundaries.

\begin{figure}
\centerline{
\psfig{figure=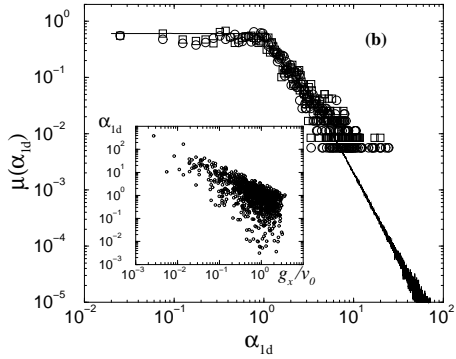,width=6cm,angle=0}
}
\caption{
Experimental histogram of $\alpha_{1d}$ for steel beads (circles) and
glass beads (squares).  Line: $\mu(\alpha_{1d})$ obtained in the RRC
model as selected by the collisional dynamics (see text).  
Inset:
Experimental scatter-plot of $\alpha_{1d}$ versus relative
precollisional horizontal velocity $g_x$, for glass beads. Data
from~\cite{Rouyer} and~\cite{Feitosa}.  }
\label{fig:rhoalpha}
\end{figure}

The horizontal component of relative velocities are computed before
($g_x$) and after ($g^*_x$) each collision, from which we deduce the
effective restitution coefficient
\begin{equation}
\alpha_{1d} = \frac{|g^{*}_x|}{|g_x|} .
\label{eq:alpha}
\end{equation}
Fig.~\ref{fig:rhoalpha} displays the histogram
$\mu_{exp}(\alpha_{1d})$ obtained from the experimental data for 
different materials.  At large $\alpha_{1d}$, a power-law tail is
evidenced. Note that values $\alpha_{1d}>1$ are expected, due to the
transfer from vertical to horizontal translational kinetic 
energy~\cite{Barrat_Trizac}.

The strong correlations between relative horizontal velocity $g_x$ and
$\alpha_{1d}$ are clearly seen in the scatter-plot (inset of
Fig.~\ref{fig:rhoalpha}), with a very sharp cutoff above the second
bisector $\alpha_{1d} \propto 1/g_x$. This cut-off follows from the
definition of $\alpha_{1d}$: since the post-collision velocity is
finite, large values of $\alpha_{1d}$ may only be obtained for small
values of $g_x$, in which case Eq.~(\ref{eq:alpha}) implies that the
maximum $\alpha_{1d}$ is of order $1/g_x$.

These features are qualitatively the same for all materials
investigated. Moreover, three different densities have been
investigated for steel beads, and the same distributions have been
obtained~\cite{Feitosa}. Similar distributions have also been measured
in a different experimental set-up with rolling beads~\cite{blair}.
The relatively small number of collisions investigated does not
however allow us to get accurate histograms for the joint
distributions $\mu(\alpha_{1d},g_x)$ nor the {\em conditional}
$\mu(\alpha_{1d}|g_x)$. The correlations evidenced in
Fig.~\ref{fig:rhoalpha} nevertheless play a crucial role,
as will be shown below.

More insight into the conditional distributions
$\mu(\alpha_{1d}|g_x)$ has been obtained in molecular dynamics of
two-dimensional IHS driven by vibrating walls in
Ref.~\cite{vibrated}. Histograms of $\mu(\alpha_{1d})$ similar to the
experimental results have been obtained. It turns out that the global
$\mu(\alpha_{1d})$ is almost insensitive to the details of the system
(density, velocity of the vibrating walls...), while this dependence 
exists for $\mu(\alpha_{1d}|g_x)$ (and also for $P(v)$). The following
characteristics of $\mu(\alpha_{1d}|g_x)$ have been obtained: at
constant $g_x$, $\mu(\alpha_{1d}|g_x)$ is almost constant for
$\alpha_{1d} \in [0,\alpha_0]$, has a small peak at
$\alpha_0$,
%
%\footnote{this results in a small peak at $\alpha_0$ for the
%global $\mu(\alpha_{1d})$; this peak is not observed in the
%experimental histograms, either for lack of statistics or because the
%experimental restitution coefficient is not really constant so that
%the peak is smoothed out, see e.g.~\cite{King_Menon} for recent
%experiments on the measure of restitution coefficients.}
and decreases as $\exp( - A(g_x) \alpha_{1d}^2 )$ for $\alpha_{1d} >
\alpha_0$, with $A(g_x) \propto g_x^2$.

\section{Random Restitution Coefficient model}
\label{sec:3}
We have considered the possibility to mimic the experimental
distributions by including randomness in the restitution coefficient,
following the approach of \cite{Barrat_Trizac} where a random restitution
coefficient (RRC) model was introduced to
account for the fact that, in vertically shaken granular gases, the
energy is transferred to the vertical degrees of freedom by the moving
piston, and then to the horizontal ones through grain/grain collisions
only.  The heating of horizontal degrees of freedom thus occurs through the
inter-particle collisions, and not in between as in the ``stochastic
thermostat'' approach. Moreover, a globally dissipative collision may
correspond to an energy gain for the horizontal components of the
velocities.  This leads to the study of horizontally projected
collisions with an effective restitution coefficient that can be
either smaller or larger than 1, as the experimental data of
Fig.~\ref{fig:rhoalpha} indeed show. In our case, the RRC model is
therefore an effective approach in $1$ dimension (since the original
collisions are two-dimensional), in which IHS undergo binary,
momentum-conserving collisions with a restitution coefficient
$\alpha_{1d}$ drawn randomly
 at each collision from a given distribution $\mu(\alpha_{1d})$
which should mimic $\mu_{exp}(\alpha_{1d})$.
Even if the original collisions are not random, the projected ones
may be considered as such.

Note that the energy injection is here given solely by the values of
$\alpha_{1d}$ larger than $1$.  The simulations are performed at the
mean-field level of the homogeneous non-linear Boltzmann equation for
point particles, solved by the Direct Simulation Monte Carlo method
(DSMC)~\cite{Bird}.  The velocity statistics is computed in the
non-equilibrium steady state which is reached after a transient.

We have first considered distributions with a large tail in order to
reproduce the experimental $\mu(\alpha_{1d})$, but without any
correlations with the relative velocities of the colliding particles;
in this case, it turns out that $P(v)$ has a power-law decay at large
$v$, in marked contrast with experimental results~\footnote{It is
noteworthy that a large $v$ power law is actually obtained
for {\em any} distribution $\mu(\alpha_{1d})$,
see~\cite{Barrat_Trizac}.}.  This approach consequently needs to be
refined and the next crucial step is to take into account the
correlations between $\alpha_{1d}$ and $g_x$, with the insight given
by the experimental scatter-plot (inset of Fig.~\ref{fig:rhoalpha})
and by the molecular dynamics results~\cite{vibrated}.

We have thus used distributions decaying as $\mu(\alpha_{1d}|\tilde
g_x) \sim \exp[-(\alpha_{1d} \tilde g_x)^2/R]$ at large $\alpha_{1d}$%
%(see Fig.~\ref{fig:rhoalphag})
, where $\tilde g_x=g_x/v_0$ is the
rescaled velocity defined from the total kinetic energy of the system
($v_0^2=\langle v^2\rangle$), and the parameter $R$ can be varied with
values of order 1. The function $\mu(\alpha_{1d} | g_x)$ is the only
input needed to simulate the RRC model. For the consistency of the
approach, the distribution $\mu(\alpha_{1d})$ measured in the
simulation needs to be close to its experimental counterpart. This
comparison is displayed in Fig.~\ref{fig:rhoalpha} and justifies {\it
a posteriori} the choice made for $\mu(\alpha_{1d} | g_x)$. Both
experimental and numerical distributions $\mu(\alpha_{1d})$ display a
power law tail of the form $\alpha_{1d}^{-n}$ with $n \simeq 3$.

The velocity distribution obtained from the RRC model is compared to
the experimental measure in Fig.~\ref{fig:pv}.  The agreement is
satisfactory over the whole range of velocities; in particular, the
RRC distributions is compatible with the stretched exponential
behavior reported in~\cite{Rouyer}, with an exponent $\nu$ close to
1.5.  Note that, since no precise experimental data is available for
the distributions of restitution parameters conditioned by relative
precollisional velocity, the parameters $R$ are tunable [as long as
the global $\mu(\alpha_{1d})$ coincides with the experimental one],
and the value giving the best agreement has been chosen ($2 \le R \le
4$). The agreement remains satisfactory upon changing $R$, provided
that the resulting large $\alpha$ cutoff remains sharp (i.e. $R$
should not be too large).

The RRC model therefore provides a self-consistent framework 
which allows to reproduce the experimental $P(v)$ if
implemented with the correct distribution of effective coefficients. 
Moreover, the velocity statistics depends on the distribution of
effective restitution coefficients: 
a broader $\mu(\alpha_{1d} | g_x)$ leads to a broader $P(v)$, consistently
with the numerical study of~\cite{vibrated} which 
showed both broader $P(v)$ and
$\mu(\alpha_{1d} | g_x)$ as e.g. the density is increased.
Both distributions $P$ and $\mu$ are equally sensitive to a possible
non universality (dependence on material properties).
As a consequence, an accurate experimental measure of $\mu$ appears as 
complementary to the direct computation of $P(v)$, in order to assess
the experimentally difficult question of the velocity statistics 
universality.

\section{Stochastic thermostat model}
\label{sec:4}
For comparison, we have also considered heating of inelastic hard
discs (2D) through the ``stochastic thermostat'', in the framework of
the non-linear homogeneous Boltzmann equation, where the
large velocity tail has been shown to behave like $\exp(-v^{3/2})$~\cite{twan}:
this result has subsequently often been considered as an agreement with the
experimental results.  In this model, the energy injection is achieved
through a random force ${\mathbf \eta}(t)$ acting on each particle
\begin{equation}
\frac{d {\mathbf v}}{d t}={\mathbf F} + 
{\mathbf \eta(t)},\;\;\;\langle \eta_i(t)\eta_j(t')\rangle 
=2D\delta(t-t')\delta_{ij}
\label{randomforces}
\end{equation}
where $D$ is the amplitude of the injected power and ${\mathbf F}$ the
systematic force due to inelastic collisions. 
The variance of $\mathbf \eta$ determines the granular temperature in the
non equilibrium steady state, but has  
no influence on the form of the rescaled distribution
function $P(c_x)$.

\begin{figure}[htb]
\centerline{
\psfig{figure=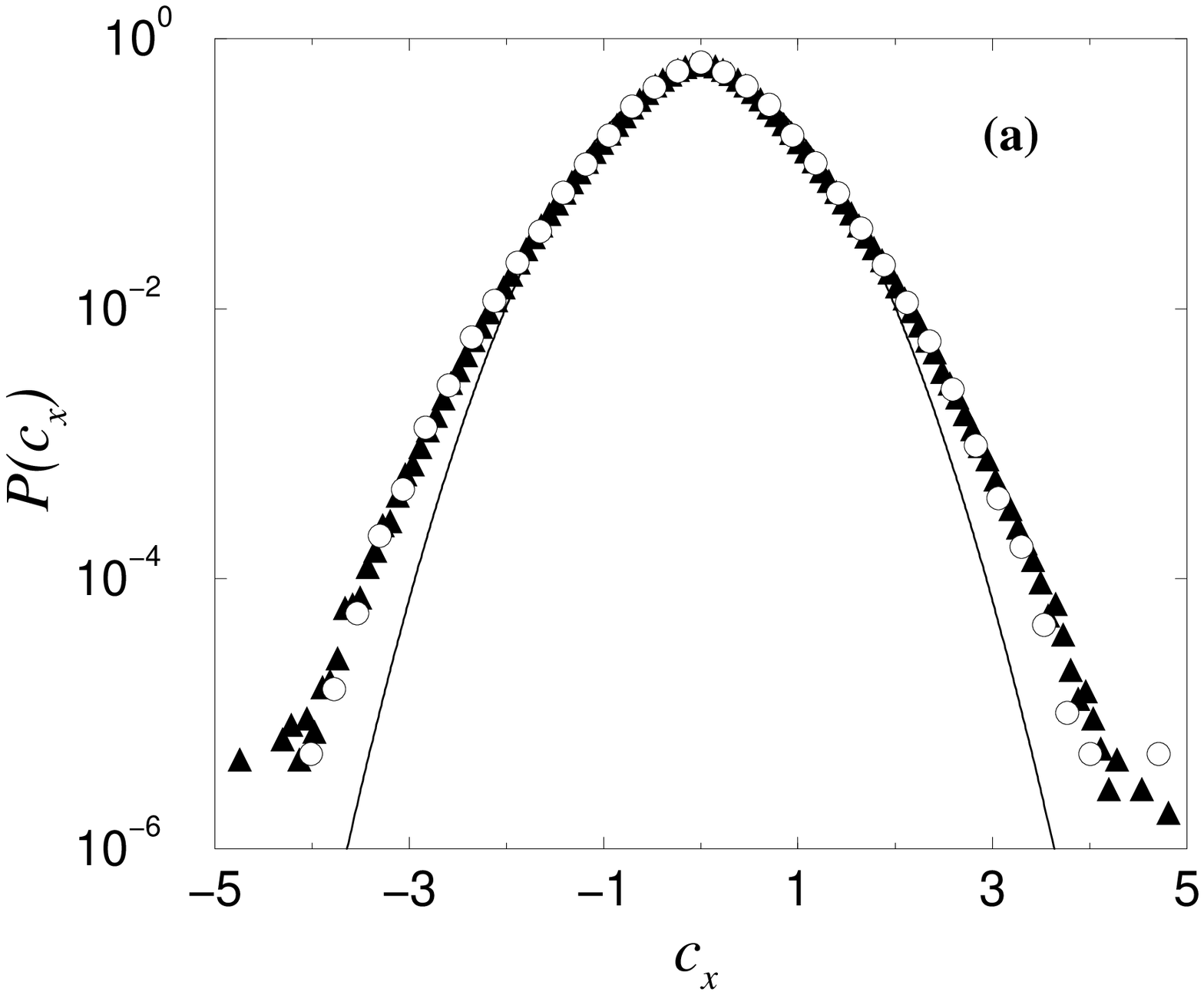,width=6cm,angle=0}
}
\centerline{
\psfig{figure=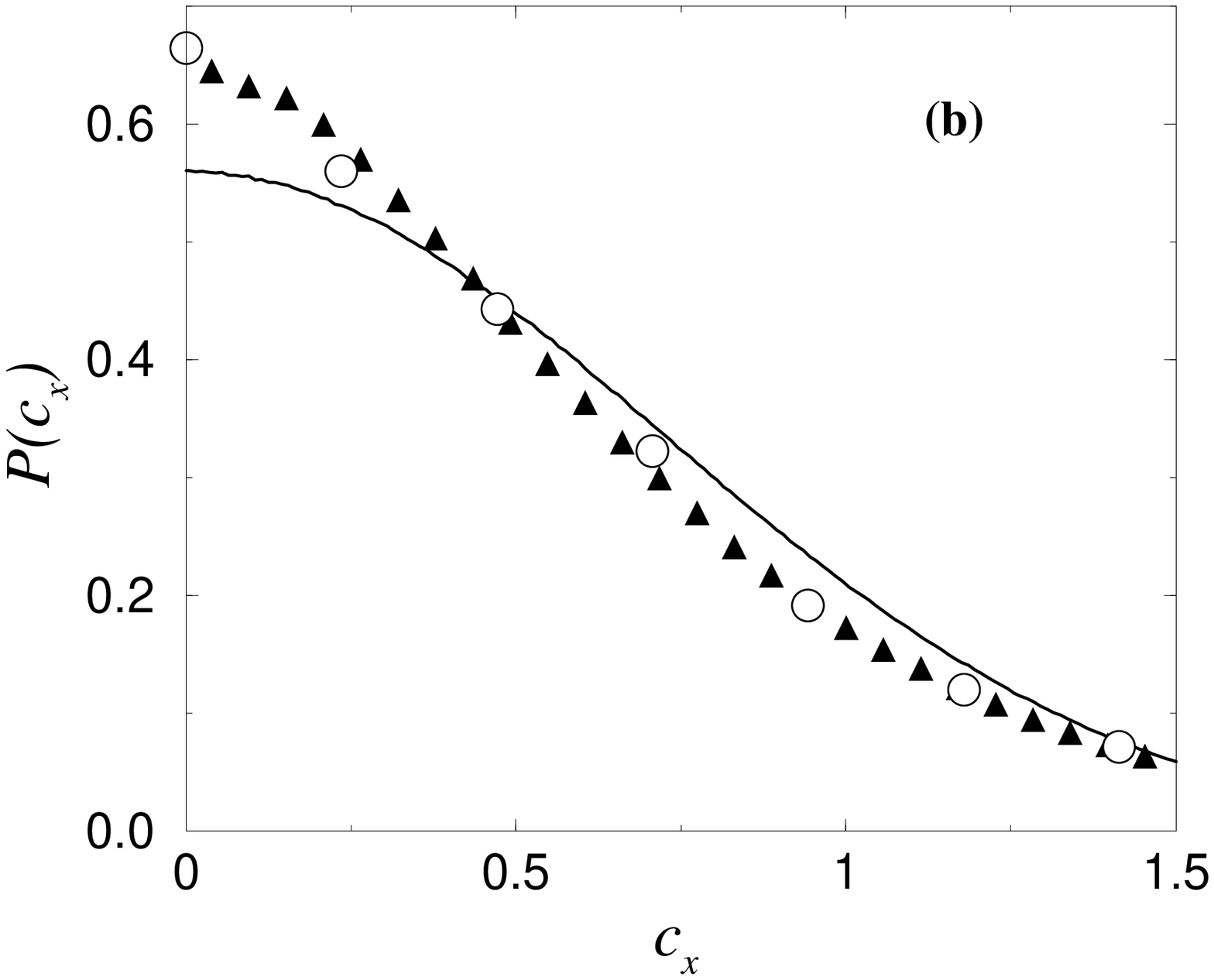,width=6cm,angle=0}
}
\caption{Rescaled distribution $P(c_x)$ of horizontal velocities, 
on a linear-log scale (a) and linear scale (b).  All distributions
have the same variance $\langle c^2_x \rangle=1/2$.  Circles represent
the experimental data for steel beads~\cite{Rouyer}; Filled triangles
correspond to the Monte Carlo simulation of the RRC model, with
$\alpha_{1d}/g_x$ correlations.  
}
\label{fig:pv}
\end{figure}

\begin{figure}[htb]
\centerline{
\psfig{figure=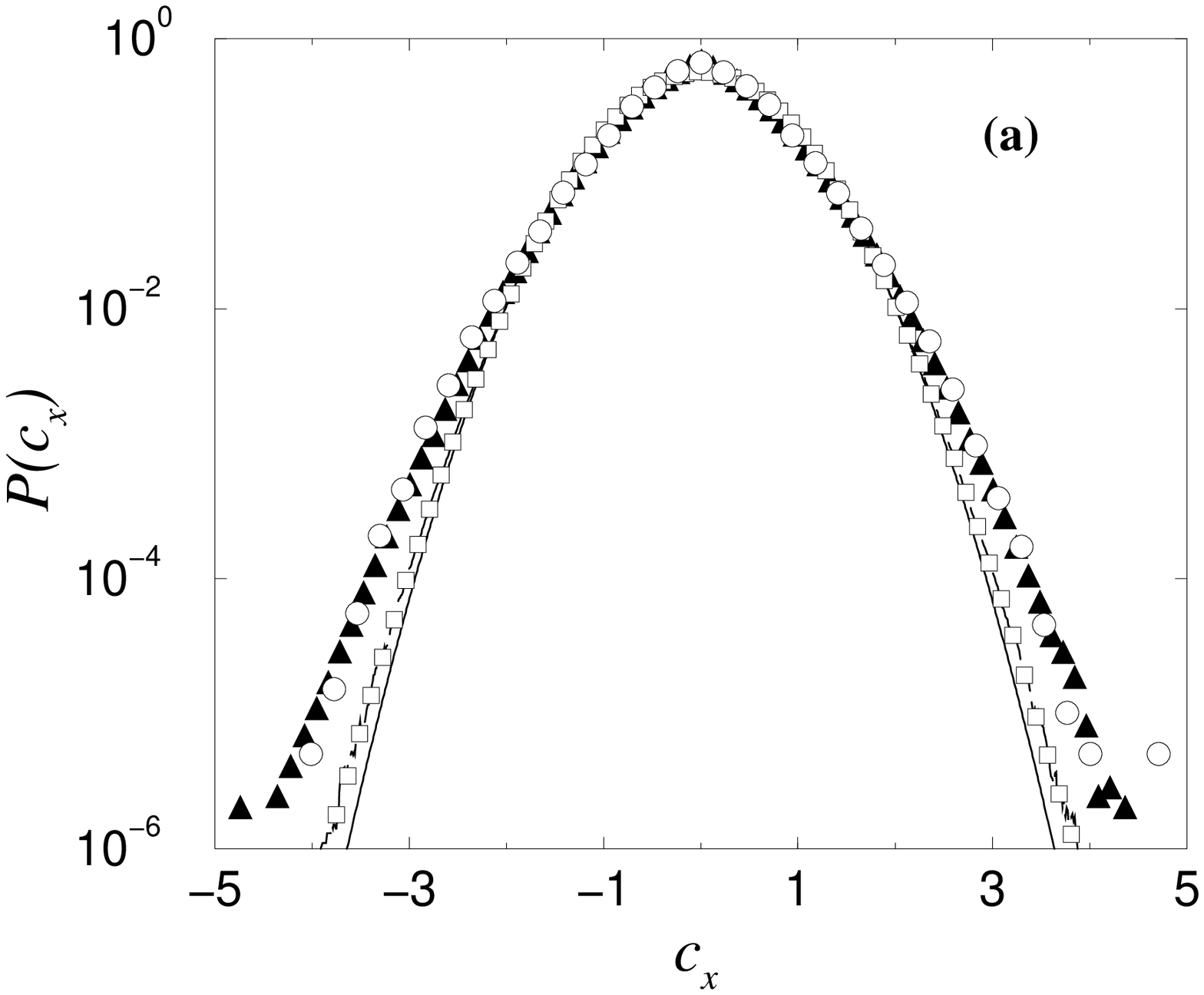,width=6cm,angle=0}
}
\centerline{
\psfig{figure=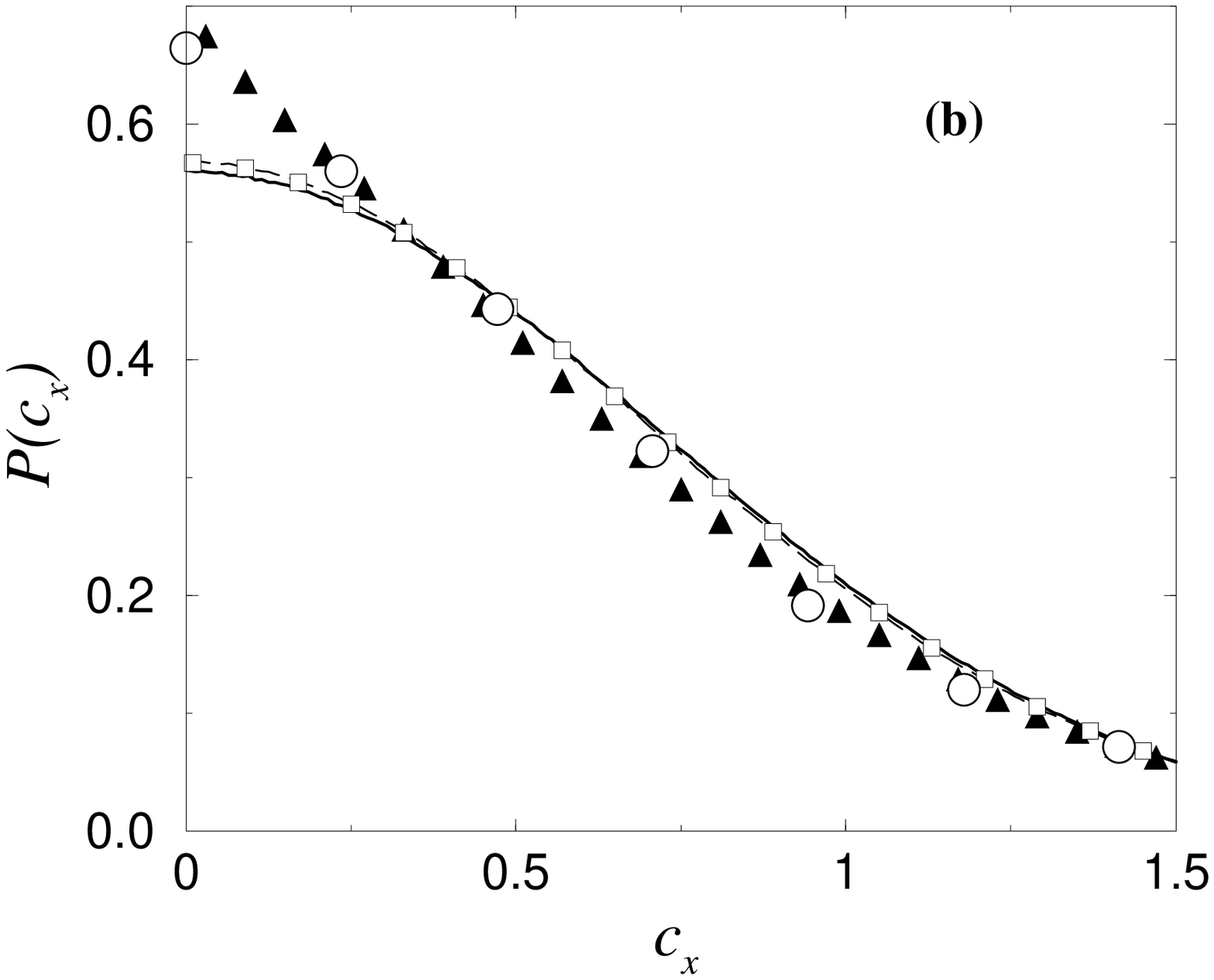,width=6cm,angle=0}
}
\caption{Rescaled distribution $P(c_x)$ of horizontal velocities,
on a linear-log scale (a) and linear scale (b).  All distributions
have the same variance $\langle c^2_x \rangle=1/2$.  Circles represent
the experimental data for steel beads~\cite{Rouyer}; 
The squares correspond to a
simulation of the ``stochastic thermostat'' with $\alpha_0 = 0.6$,
whereas for $\alpha_0 > 0.7$, the corresponding $P(c_x)$ is
indistinguishable from the Gaussian shown by the full line.  
The dashed line (very close to the squares) 
corresponds to a simulation of the three parameters
model~\cite{Foerster,Walton,Luding_pre} 
with $\alpha_n=0.7$, $\alpha_t=0.5$, $\mu=0.5$.
Filled triangles show the results for the multiplicative driving
stochastic thermostat with $\delta=0.6$ and
$\alpha=0.9$~\cite{Cafiero_prl}.
}
\label{fig:pv_new}
\end{figure}

With the accuracy of Fig.~\ref{fig:pv_new} (the current experimental
resolution), the corresponding numerical velocity distributions are
then found indistinguishable from a Gaussian, for physically relevant
inelasticities in the range $0.7 \leq \alpha_0 \leq 1$.  Departure
from Maxwell-Boltzmann behavior becomes manifest below $\alpha_0 =
0.6$ (squares in Fig.~\ref{fig:pv_new}), which is unphysically low, but
the velocity distribution is still incompatible with its
experimental counterpart.  We also investigated the
possibility to describe the effective horizontal dynamics with the 1D
stochastic thermostat: for $0.7 <\alpha_0<1$ the velocity
distributions are incompatible with the experimental $P(v)$ displayed
in Figs.~\ref{fig:pv}, with opposite non Maxwellian features
(underpopulated both at vanishing and high energies~\cite{BBRTW}).
More precisely, within the stochastic thermostat approach the kurtosis
$\langle v_x^4 \rangle/\langle v_x^2\rangle^2-3$ of the distribution
is negative for $\alpha_0 > 1/\sqrt{2} \simeq 0.71$~\cite{twan},
irrespective of dimension, which corresponds to an underpopulated low
velocity behavior at variance with the experimental data shown in
Fig.~\ref{fig:pv}.

We have also considered the stochastic thermostat mo\-del for two-dimensional
IHS with both tangential $\alpha_t$ and normal $\alpha_n (=\alpha_0)$
restitution coefficients \cite{Luding-McNamara}.  No numerical studies of
$P(v)$ can indeed be found in the literature in this case, although an
investigation into the non-equipartition between translational and rotational
kinetic energies has been performed in~\cite{Luding-McNamara}.  The resulting
velocity distributions $P(v)$ remain very close to a Gaussian for $\alpha_n
\ge 0.7$ and arbitrary $\alpha_t$ (where $-1\leq \alpha_t\leq 1$), as for
smooth spheres (corresponding to $\alpha_t=-1$). However, such a two parameter
model may be too schematic compared to the experiments~\cite{Foerster} 
and we have 
also considered a more realistic approach with Coulomb friction along the
simplifications discussed in~\cite{Walton}: a friction coefficient 
$\mu$, is introduced
in addition to ($\alpha_t,\alpha_n$)~\cite{Luding_pre}.
This does not change significantly $P(c_x)$
(see the squares and dashed line in Fig.~\ref{fig:pv_new}).
This seems to discard the relevance
of such an approach for the comparison of the velocity distributions with
experimental data.

The above analysis shows that the stochastic thermostat in its original
formulation (including some possible extensions) 
does not provide a relevant model of energy injection
as far as the velocity distribution is concerned,
although it may be useful to investigate other features such as kinetic
energy non-equipartition in granular mixtures \cite{Granular}.
However, a variant of this model may improve the picture. 
In particular, a multiplicative driving [corresponding to a velocity
dependent amplitude
$D \propto |{\bf v}|^{2 \delta}$  in Eq. (\ref{randomforces})]
has been studied in Refs. \cite{Cafiero_prl,Cafiero}. We have performed 
DSMC simulations for this model, choosing the value of $\delta$ 
that, for a given reasonable inelasticity parameter ($\alpha=0.9$),
gives the best agreement with the experimental $P(v)$. We obtained 
$\delta \simeq 0.6$, which leads to the distribution shown by the triangles 
in Fig. \ref{fig:pv_new}. The agreement with the experimental data
is satisfactory, and displays a similar accuracy as obtained within
the RRC model. Consequently, models with homogeneous energy
injection may also describe quite accurately the experimental $P(v)$,
with the problem of predicting the values of the various parameters
involved.

\section{Conclusion}
In conclusion, the Random Restitution Coefficient model with point
particles captures the essential features responsible for the observed
non Gaussian character of the velocity distribution $P(v)$ in vibrated
granular gases experiments~\cite{Rouyer}, and represents therefore a
self-consistent framework. The
conditional distribution $\mu(\alpha_{1d}|g_x)$ defining the
appropriate collision rule has been shown to encode the relevant
dynamic information and provides an alternative route to characterize
the non equilibrium steady state, complementary to the direct measure
of $P(v)$. An interesting point would be to obtain an analytical
prediction for $\mu(\alpha_{1d}|g)$.  It is noteworthy that our
approach is mean-field (Boltzmann) like, the only correlations
considered being in the collision law. Its self-consistency, 
which was not an
obvious point {\it a priori}, has been established by comparison with
experiments.

While we have restricted our analysis to the two-dimensional case,
such investigations can be extended to three-dimensional
systems, for which however experimental measures of effective
restitution coefficients seem more difficult.
As implemented here, without an analytical knowledge of
$\mu(\alpha_{1d}|g_x)$, the RRC model is not predictive
since an experimental input is required to
obtain the correct velocity statistics. 
Our results however suggest to assess experimentally the
question of the universality of $P(v)$ from the direct measure of the
distribution of effective restitution coefficients. These
characteristics are indeed linked within the RRC
model: the exponent $\nu$, and the whole shape of $P(v)$, depend 
on the functional form of $\mu(\alpha_{1d}|g_x)$.  
At this point, the fact that with a rather poor accuracy, similar 
$P$ and $\mu$ have been obtained for the case investigated in 
\cite{Rouyer,Feitosa} simply confirms the RRC picture, and calls 
for experiments with widely different collisional properties,
such as hollow spheres.

Acknowledgments:
We are grateful to K. Feitosa and N. Menon for generous provision of
unpublished experimental data and interesting correspondence.

\end{document}